\documentclass[lettersize,journal]{IEEEtran}
\usepackage{amsmath,amsfonts}
\usepackage{algorithmic}
\usepackage{algorithm}
\usepackage{array}
\usepackage[caption=false,font=normalsize,labelfont=sf,textfont=sf]{subfig}
\usepackage{textcomp}
\usepackage{url}
\usepackage{verbatim}
\usepackage{graphicx}
\usepackage{cite}

\usepackage{booktabs}	% manual add
\usepackage{multirow}	% manual add
\usepackage{tabularx}   % manual add
\usepackage{makecell}   % manual add
\usepackage{subfig}		% manual add
\begin{document}
\title{MAD: A Multimodal and Multi-perspective Affective Dataset with Hierarchical Annotations}
\author{Shengwei~Guo, Yunqing~Qiao, Wenzhan~Zhang, Bo~Liu, Yong~Wang, and Guobing~Sun
		\thanks{Corresponding author: Guobing Sun (e-mail: sunguobing@hlju.edu.cn).}}
\markboth{Journal of \LaTeX\ Class Files,~Vol.~14, No.~8, August~2021}%
{Shell \MakeLowercase{\textit{et al.}}: A Sample Article Using IEEEtran.cls for IEEE Journals}

\IEEEpubid{0000--0000/00\$00.00~\copyright~2021 IEEE}

\maketitle

\begin{abstract}
This work presents MAD (Multimodal Affection Dataset), a multimodal emotion dataset designed for affective computing and neurophysiological modeling. MAD is built upon synchronous collection of diverse physiological signals (EEG, ECG, EOG, EMG, PPG, and BCG) together with tri-view RGB-D facial videos, enabling the observation of emotional dynamics from neural, physiological, and behavioral perspectives.

The dataset consists of synchronized recordings from 18 participants and introduces two key contributions. First, it provides temporally aligned multimodal data that jointly capture central neural activity, peripheral physiological responses, and overt facial expressions. Second, it incorporates a three-level emotion annotation framework spanning stimulus elicitation, subjective cognition, and behavioral expression, supporting joint modeling of the full emotion process.

To validate the dataset, we conduct systematic benchmark experiments covering intra-subject EEG emotion recognition, cross-subject EEG transfer learning, consistency analysis and emotion classification with cardiac-related signals, multimodal physiological fusion, and multi-view facial emotion recognition. The experimental results demonstrate that MAD supports consistent and comparable performance across both unimodal and multimodal settings, establishing it as a reliable benchmark for emotion recognition and cross-modal affective analysis, and as a valuable resource for studying emotion mechanisms across multiple levels.
\end{abstract}


\section{Introduction}

\label{sec:introduction}
Emotion recognition is a core task in affective computing and plays a critical role in psychological health monitoring, clinical decision support, human–computer interaction, and intelligent systems~\cite{romero2025positive, gu2025prefrontal}. Existing approaches primarily rely on observable behavioral modalities such as text, speech, or facial expressions, which are easy to acquire and naturally expressive, and thus perform well under controlled conditionss~\cite{pillalamarri2025review, behzad2025facial}. However, overt behaviors represent the final output of emotional processes and are highly susceptible to individual expression habits, social context, attentional regulation, and even deliberate concealment\cite{das2023multimodal, li2020deep, kollias2019deep}. As a result, their surface-level features do not always faithfully reflect an individual’s internal emotional state, nor do they adequately reveal the underlying neural processing and autonomic physiological regulation involved in emotion generation\cite{schlicher2025emotionally, d201431, pei2024affective}.

In contrast, physiological signals characterize emotion-related neural and autonomic regulation processes with minimal subjective intervention, offering advantages such as higher objectivity and stronger resistance to intentional manipulation\cite{afzal2024comprehensive, gross2015emotion, barrett2017theory}. Nevertheless, a single physiological modality can only capture a partial aspect of the emotional process. For example, EEG reflects neural activity but is sensitive to artifacts such as body movements\cite{uriguen2015eeg}; cardiac-related signals such as ECG can indicate emotional arousal levels but have limited discriminative power for fine-grained emotion categories; EOG and EMG are more closely associated with behavioral responses than with emotional states themselves\cite{kreibig2010autonomic,kunecke2014facial}. Consequently, unimodal approaches are insufficient to support comprehensive modeling of complex emotional states.

From psychological and neuroscientific perspectives, emotion is a multi-stage, cross-level process encompassing stimulus elicitation, cognitive appraisal, and overt behavioral expression. Information across these levels may be consistent or may diverge substantially\cite{scherer1982emotion, xue2021physiological}; for instance, external stimuli, internal subjective experiences, and final facial expressions are often asynchronous\cite{ahmad2022survey,critchley2013visceral} . To systematically investigate the complete “elicitation–cognition–expression” emotional process, it is necessary to simultaneously observe multiple physiological modalities and overt visual behaviors, while establishing a multi-level emotion annotation framework.

However, existing public datasets (e.g., FER2013~\cite{goodfellow2013challenges}, DEAP~\cite{koelstra2011deap}, SEED~\cite{zheng2015investigating}, and MAHNOB-HCI~\cite{miranda2018amigos}) exhibit notable limitations in supporting such cross-level investigations. First, most datasets provide only a single source of emotion labels (either stimulus-based or cognitive), making it difficult to analyze consistency and discrepancies across different levels. Second, the coverage of physiological modalities is limited, with most datasets including only EEG or a small number of peripheral signals, which constrains the study of neural–peripheral complementarity and hinders exploration of more accessible alternative modalities such as PPG, BCG, and rPPG for affective recognition. Third, visual data are typically restricted to single-view 2D RGB videos, limiting research on cross-view robustness and three-dimensional structural perception.

To address these limitations, we introduce the \textbf{Multimodal Affection Dataset (MAD)}, a multimodal emotion dataset designed to support cross-level and cross-modal investigations of emotional processes. By jointly recording diverse physiological signals together with multi-view facial observations and organizing emotion annotations across elicitation, cognition, and expression, MAD enables systematic analysis of emotional consistency, complementarity, and variability across neural, physiological, and behavioral levels. 
﻿ 
The primary contributions of this work are summarized as follows: 
﻿ 
\begin{itemize} 
	\item \textbf{A cross-level multimodal emotion dataset:} MAD integrates central and peripheral physiological signals with multi-view facial behavior, providing a unified data foundation for studying how emotional responses evolve from neural and autonomic regulation to observable expression. 
	﻿ 
	\item \textbf{A hierarchical emotion annotation framework:} We propose a three-level labeling scheme covering stimulus elicitation, subjective cognition, and behavioral expression, enabling quantitative comparison of emotional information across different conceptual layers.
		﻿ 
	\item \textbf{Benchmark protocols for multi-level emotion analysis:} We establish a set of representative benchmark tasks spanning intra-subject and cross-subject EEG recognition, cardiac-related signal analysis, multimodal physiological fusion, and multi-view facial expression recognition. These benchmarks demonstrate the applicability of MAD for evaluating emotion models across modalities, subjects, and observation perspectives. 
\end{itemize} 
﻿ 
Overall, MAD provides a unified experimental platform for multi-level emotion modeling and cross-modal affective analysis, facilitating future research on emotion mechanisms, representation learning, and interpretability across neural, physiological, and behavioral domains. 
﻿

\begin{figure*}[!t]
	\centerline{\includegraphics[width=.78\linewidth]{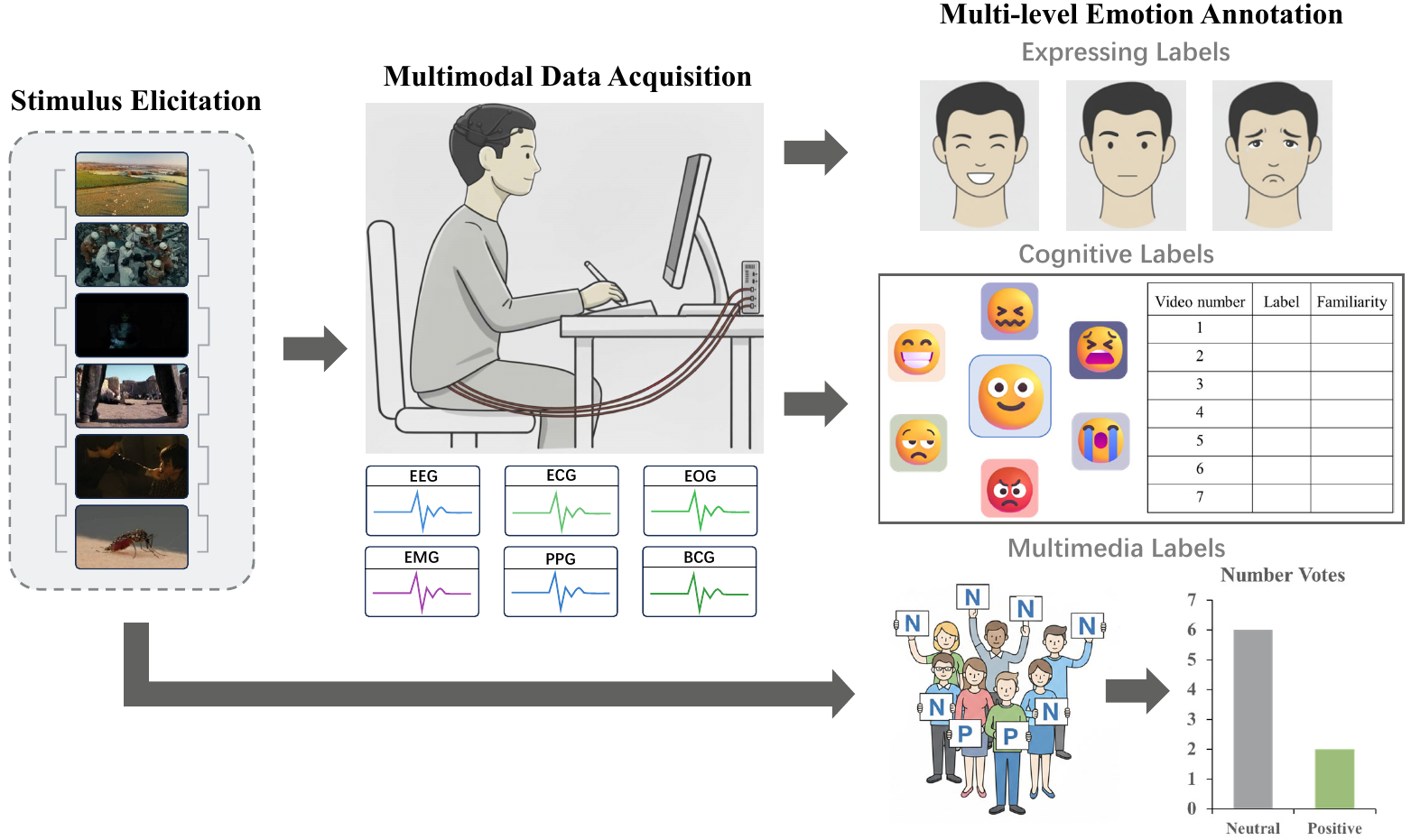}}
	\caption{Overview of the experimental pipeline. 
		The left panel illustrates the emotion elicitation stage based on video stimuli (\textit{Stimulus Elicitation}), 
		the middle panel shows the synchronized acquisition of multimodal physiological and visual signals during stimulus presentation (\textit{Multimodal Data Acquisition}), 
		and the right panel depicts the hierarchical emotion annotation process, including stimulus-based, self-reported, and expression-based labels (\textit{Multi-level Emotion Annotation}).}
	\label{fig:collect_data}
\end{figure*}

\section{Related Work}
In emotion recognition research, visual-based methods, physiological signal–based approaches, and multimodal fusion techniques have all achieved substantial progress~\cite{jafari2023emotion}. Existing studies are generally categorized into three groups: vision-based emotion recognition focusing on behavioral expression analysis, physiology-based emotion recognition targeting internal state analysis, and comprehensive modeling approaches that integrate multiple information sources. Despite continued advances in each direction, publicly available datasets still exhibit notable limitations in modality coverage, annotation hierarchy, and scale.

\subsection{Vision-Based Emotion Recognition}

Facial expressions constitute one of the most intuitive and easily accessible overt behavioral cues for emotion expression~\cite{liu2023brain, liu2017real, liu2019ntu}. Classic datasets such as FER2013~\cite{goodfellow2013challenges} and FER+~\cite{barsoum2016training} provide large-scale samples with basic emotion labels, while Multi-PIE~\cite{GROSS2010807} and RaFD~\cite{langner2010presentation} further introduce variations in pose and illumination to enhance model generalization~\cite{huang2014novel, soleymani2011multimodal}.

However, these datasets predominantly rely on single-view 2D RGB images and lack depth information as well as multi-view representations, making it difficult for models to learn stable three-dimensional emotional representations. As a result, they exhibit clear limitations in cross-view recognition, robustness to occlusion, and deployment in real-world scenarios.

\subsection{Physiological Signal–Based Emotion Recognition}

Compared with visual modalities, physiological signals can more directly reflect autonomic nervous system activity and are less influenced by conscious control, thereby offering distinct advantages for emotion recognition. Signals such as EEG, ECG, EOG, and EMG have been widely employed in affective computing research~\cite{wang2022multi}.

Representative datasets such as SEED~\cite{zheng2015investigating} and SEED-IV~\cite{zheng2018emotionmeter} systematically demonstrate the discriminative capability of EEG features in classifying positive, neutral, and negative emotions, with SEED-IV further incorporating eye-tracking signals to support finer-grained emotion modeling. In addition, MAHNOB-HCI~\cite{miranda2018amigos} and DEAP~\cite{koelstra2011deap} introduce multiple physiological modalities combined with subjective self-assessment labels, providing early foundations for multimodal emotion analysis.

Nevertheless, these datasets are primarily designed to validate the effectiveness of individual or limited modalities, and thus remain insufficient in terms of signal modality coverage, joint modeling across physiological systems, and hierarchical emotion annotation structures. In particular, most datasets do not simultaneously support systematic investigation of consistency, complementarity, and functional roles among multiple physiological signals across different emotional levels, thereby limiting holistic analysis of the physiological foundations of emotion.

\subsection{Multimodal Fusion and Emerging Physiological Signals}

The fusion of visual and physiological modalities has become an important direction for enhancing the robustness and generalizability of emotion recognition systems~\cite{zhang2020emotion1, zhang2020emotion2, liu2023emotionkd}. With advances in wearable sensing and depth imaging technologies, multimodal emotion analysis is gradually extending beyond controlled laboratory settings toward more naturalistic and real-world scenarios.

In parallel, emerging physiological signals such as photoplethysmography (PPG) and ballistocardiography (BCG) have attracted increasing interest. These signals capture cardiovascular dynamics through optical or mechanical measurements and provide complementary information to conventional modalities such as EEG and ECG. Recent studies have demonstrated their potential for affective analysis, particularly in non-invasive or low-contact settings~\cite{schmidt2018introducing, liao2022comparison, habib2022performance}.

Despite these advances, existing studies are often conducted on heterogeneous or partially synchronized data, and are typically limited to a narrow subset of modalities or observation perspectives. As a result, the interplay between emerging cardiac-related signals, central neural activity, and visual emotion expression remains insufficiently explored at the dataset level.

\subsection{Limitations of Existing Datasets and Positioning of MAD}

A closer examination of existing public emotion datasets reveals several common limitations that hinder comprehensive multimodal and cross-level emotion analysis.

First, modality coverage is often restricted. Most datasets include only a single central modality (e.g., EEG) or a small number of peripheral signals, and rarely support joint analysis across neural, cardiovascular, and behavioral systems. This limitation constrains the study of modality complementarity and heart--brain interactions.

Second, emotion annotations are typically confined to a single conceptual level, such as stimulus-based or self-reported labels. The absence of multi-level annotations spanning emotion elicitation, subjective cognition, and behavioral expression makes it difficult to investigate consistency and divergence across different stages of emotional processing.

Third, visual data are commonly limited to single-view 2D RGB recordings, which restricts research on cross-view robustness and three-dimensional facial expression modeling in realistic scenarios.

To address these gaps, MAD (Multimodal Affection Dataset) is designed as a unified experimental resource that integrates diverse physiological modalities with multi-view visual observations and multi-level emotion annotations, providing a data foundation for cross-modal, cross-level emotion research.

\section{Dataset Design and Acquisition}

This section describes the design principles, experimental protocol, multimodal acquisition devices, annotation framework, and basic preprocessing procedures of the MAD dataset. The purpose of this section is to ensure the reproducibility and transparency of data acquisition, thereby providing a solid data foundation for subsequent modeling and benchmark experiments.

\subsection{Subjects and Experimental Protocol}

A total of 18 healthy participants were recruited for this study (9 males and 9 females, aged 23–25 years). All participants had normal hearing and vision, reported no history of neurological disorders, and signed informed consent forms in accordance with the institutional ethical review protocol. To minimize external influences, participants were instructed to avoid caffeine intake and strenuous physical activity for 12 hours prior to the experiment.

The experiments were conducted in a quiet and enclosed environment. Each participant watched 16 emotion-eliciting video clips with durations ranging from 1 to 7 minutes, with a 60-second resting interval inserted between consecutive clips. After each viewing session, participants completed a self-assessment questionnaire to report their emotional states. To reduce fatigue effects, the experiment was conducted over two separate days.

The overall experimental framework integrates three tightly coupled components,
namely stimulus elicitation (emotion-inducing video stimuli),
multimodal data acquisition (synchronous recording of physiological signals and facial behaviors),
and multi-level emotion annotation (including stimulus-based, cognitive, and expression-based labels),
as conceptually illustrated in Fig.~\ref{fig:collect_data}.

Specifically, the experimental protocol involves:
\begin{itemize}
	\item \textbf{Stimulus elicitation:} Emotion-eliciting video clips were presented to participants.
	\item \textbf{Multimodal data acquisition:} Physiological signals and facial behaviors were synchronously recorded while participants viewed the stimuli.
	\item \textbf{Multi-level emotion annotation:} Emotional states were annotated from stimulus-based labels, subjective self-reports, and observer-based facial expression labels.
\end{itemize}

\subsection{Emotion Stimuli and Three-Level Annotation Framework}

\subsubsection{Emotion Video Stimulus Design}

Sixteen emotion-eliciting video clips were ultimately selected from an initial pool of 50 candidates based on ratings provided by 50 independent annotators. The selected clips covered seven emotion categories: happy, angry, sad, fearful, anxiety, disgusted, and neutral.

To support emotion recognition tasks at different levels of abstraction, the stimulus-based labeling scheme is organized into three levels of granularity:

\begin{itemize}
	\item \textbf{Seven-class:} Happy, Angry, Sad, Fearful, Anxiety, Disgusted, Neutral
	\item \textbf{Four-class:} Happy, Fearful, Sad, Neutral
	\item \textbf{Three-class:} Positive, Negative, Neutral
\end{itemize}

The four-class and three-class settings provide coarse-grained affective representations that broadly align with the valence--arousal model, while maintaining a discrete classification formulation. All labels are used as stimulus-based annotations, establishing a standardized benchmark for emotion elicitation (see Table~\ref{tab:Multimedia-Labels}).

\begin{table*}[t]
	\centering
	\caption{Mapping of Multimedia Labels for the selected movie excerpts. Each clip was annotated with seven basic emotions and subsequently mapped into four‐class (happy, fearful, sad, neutral) and three‐class (positive, negative, neutral) categories.}
	\begin{tabular}{cccccc}
		\toprule
		Movie	&	Time Excerpt	&	7-class Label	&	4-class Label	&	3-class Label	\\
		\midrule
		\multirow{2}{*}{Too Cool To Kill}	&	7:30$\sim$9:14	&	Happy	&	Happy	&	Positive	\\
		&	1:36:58$\sim$1:40:35	&	Happy	&	Happy	&	Positive	\\
		\cmidrule(lr){1-5}
		\multirow{2}{*}{Welcome To Sha-ma Town}	&	1:29:02$\sim$1:29:58	&	Happy	&	Happy	&	Positive	\\
		&	1:30:20$\sim$1:32:06	&	Happy	&	Happy	&	Positive	\\
		\cmidrule(lr){1-5}
		\multirow{2}{*}{Silenced}	&	18:40$\sim$20:25	&	Angry	&	Fearful	&	Negative	\\
		&	44:29$\sim$47:15	&	Angry	&	Fearful	&	Negative	\\
		\cmidrule(lr){1-5}
		\multirow{3}{*}{Aftershock}	&	33:08$\sim$39:15	&	Sad	&	Sad	&	Negative	\\
		&	1:48:24$\sim$1:52:20	&	Sad	&	Sad	&	Negative	\\
		&	1:53:08$\sim$2:01:20	&	Sad	&	Sad	&	Negative	\\
		\cmidrule(lr){1-5}
		\multirow{2}{*}{Annabelle}	&	27:50$\sim$29:34	&	Fearful	&	Fearful	&	Negative	\\
		&	48:25$\sim$51:52	&	Anxious	&	Fearful	&	Negative	\\
		\cmidrule(lr){1-5}
		Mosquito (Doc.)	&	A fragment	&	Disgusted	&	Sad	&	Negative	\\
		\cmidrule(lr){1-5}
		Landscape (Doc.)	&	Four fragments	&	Neutral	&	Neutral	&	Neutral	\\
		\bottomrule
	\end{tabular}
	\label{tab:Multimedia-Labels}
\end{table*}
	
\subsection{Multimodal Acquisition System and Hardware Configuration}

MAD synchronously acquires six types of physiological signals and tri-view RGB-D videos through a unified timestamp system. All devices were calibrated prior to the experiments.

\begin{itemize}
	\item \textbf{EEG:} 14-channel Emotiv Epoc+, sampled at 128 Hz.
	\item \textbf{EOG/ECG/EMG:} BIOPAC MP150 system, including 1 EOG, 2 ECG, and 3 EMG channels, sampled at 1000 Hz.
	\item \textbf{PPG/BCG:} Custom wearable dual-channel PPG and seat-based BCG sensors, sampled at 34 Hz.
	\item \textbf{RGB-D:} Three Intel RealSense D455 cameras (left, frontal, and right views), recording at 640×480 resolution and 30 fps.
\end{itemize}

\begin{figure}[!t]
	\centerline{\includegraphics[width=.85\linewidth]{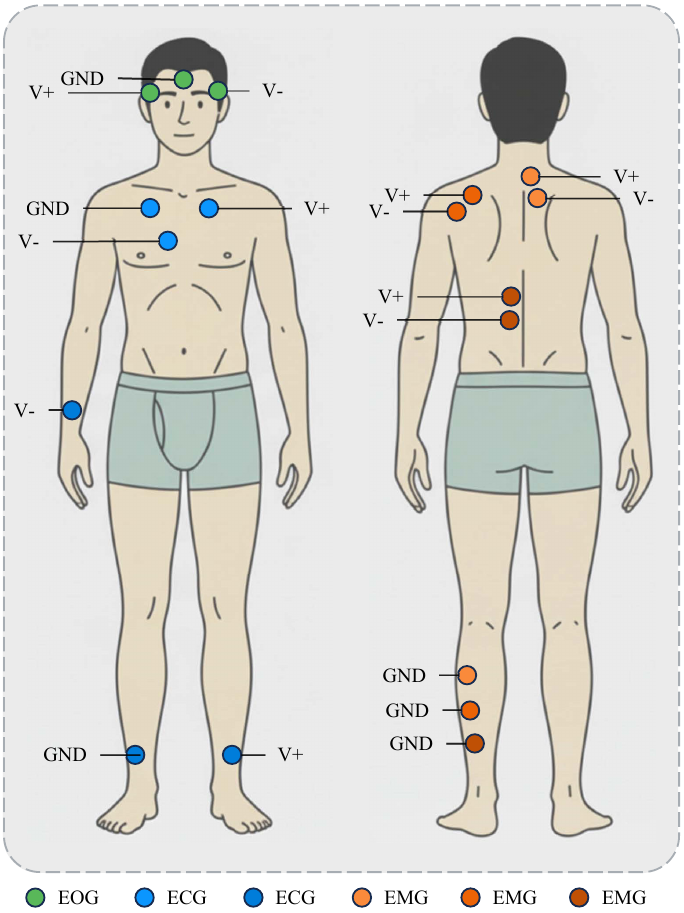}}
	\caption{Schematic illustration of electrode placement for multimodal physiological recordings. Six channels were recorded using the BIOPAC MP150 system, including one EOG, two ECG, and three EMG channels, with sensor locations indicated on the front and back schematic views.}
	\label{fig:mp150-channel}
\end{figure}

The configuration of the BIOPAC MP150 system is illustrated in Fig.~\ref{fig:mp150-channel}.

\subsection{Annotation Framework}

MAD introduces a three-level emotion annotation framework to jointly represent emotional expressions at the stimulus, cognitive, and behavioral levels:

\begin{itemize}
	\item \textbf{Stimulus labels (Multimedia):} Assigned by 50 external annotators based on video content.
	\item \textbf{Cognitive labels (Cognitive):} Seven emotion categories and intensity ratings reported by participants after viewing.
	\item \textbf{Expression labels (Expression):} Three-class emotion annotations assigned by 10 observers based on tri-view facial frames.
\end{itemize}

This framework supports systematic cross-level investigation spanning stimulus elicitation, subjective cognition, and behavioral expression.

\subsection{Dataset Scale and Basic Preprocessing}
\label{subsec:method_preprocess}

The MAD dataset occupies approximately 8 TB of storage and consists of 288 sessions in total, corresponding to 18 participants × 16 video clips.

Basic preprocessing procedures include:

\begin{itemize}
	\item All modalities were segmented into unified 60-second emotional clips corresponding to the peak periods of each stimulus;
	\item EEG signals were processed with band-pass filtering and ICA-based artifact removal;
	\item ECG, EOG, and EMG signals were resampled to 256 Hz and filtered according to their spectral characteristics;
	\item Peak synchronization between PPG/BCG and ECG signals was verified;
	\item RGB videos underwent face detection, ROI cropping, and expression frame selection.
\end{itemize}

The preprocessing steps described in this section aim to construct usable data, while specific modeling features are introduced in Section~\ref{sec:exp}.

\section{Benchmark Experiments}
\label{sec:exp}
The benchmark experiments of MAD are designed to systematically examine five core scientific questions in emotion modeling:

\begin{itemize}
	\item \textbf{(1) How sensitive is EEG to different types of emotion annotations?}
	We compare intra-subject EEG recognition performance under stimulus-based labels and cognitive labels to investigate neural-response differences between externally elicited emotions and subjective experiences.
	
	\item \textbf{(2) Can emotion models transfer effectively across individuals?}
	We evaluate cross-subject EEG emotion recognition using multiple domain adaptation methods to verify whether MAD exhibits sufficient inter-subject consistency.
	
	\item \textbf{(3) Do cardiac-related modalities (ECG, PPG, and BCG) exhibit stable consistency and potential substitutability?}
	We analyze the rhythmic structures of these three synchronized signals and compare their emotion-recognition performance to assess the feasibility of future non-contact or weak-contact acquisition.
	
	\item \textbf{(4) Can multimodal physiological fusion improve the robustness of emotion recognition?}
	We fuse EEG with ECG, EOG, and EMG to explore the complementarity across physiological systems and their contributions to performance gains.
	
	\item \textbf{(5) Can multi-view facial information improve pose-robust emotion recognition?}
	Using tri-view RGB-D videos together with contrastive learning, we study the consistency and alignability of facial expressions across views.
\end{itemize}

To address these questions, we conduct five categories of benchmark experiments: intra-subject EEG recognition, cross-subject transfer learning, consistency analysis of cardiac-related signals, multimodal physiological emotion recognition, and multi-view facial emotion recognition. All experiments follow the unified preprocessing pipeline and feature extraction procedures described in Section~\ref{subsec:method_preprocess}.

It should be noted that the sources of emotion annotations are not identical across tasks. In the intra-subject EEG experiments, we use both stimulus-based labels and cognitive labels to compare neural-response differences between externally elicited emotions and subjective experiences. In contrast, cross-subject EEG transfer, cardiac-signal analysis, and multimodal physiological fusion experiments consistently adopt stimulus-based labels as supervision to avoid confounding effects from inter-individual variations in self-assessment, thereby ensuring comparability and stability across subjects and modalities. For the multi-view facial experiments, we use manually annotated expression labels to evaluate the consistency of overt emotional expressions across different viewpoints.

\subsection{Feature Extraction and Experimental Settings}

MAD provides six synchronized physiological modalities (EEG, ECG, EOG, EMG, PPG, and BCG) as well as tri-view RGB-D videos. To ensure fair comparisons across modalities, we extract a unified set of representative time-domain, frequency-domain, and nonlinear features from each signal type, including:

\begin{itemize}
	\item EEG: power spectral density (PSD), differential entropy (DE), and Hjorth parameters (five frequency bands);W
	\item ECG: HRV features such as RR intervals, SDRR, SD1/SD2, and CVI;
	\item EMG: wavelet energy, variance, and Shannon entropy;
	\item EOG: mean, variance, PSD, and DE;
	\item PPG/BCG: HRV features consistent with those extracted from ECG.
\end{itemize}

Multimodal physiological fusion is implemented via feature-level concatenation after z-score normalization. For the visual modality, we use RGB images only, while depth frames are retained for subsequent calibration and potential rPPG studies. The visual experiments in this section are conducted on the multi-view facial expression subset of MAD.

\begin{figure*}[!t]
	\centerline{\includegraphics[width=.9\linewidth]{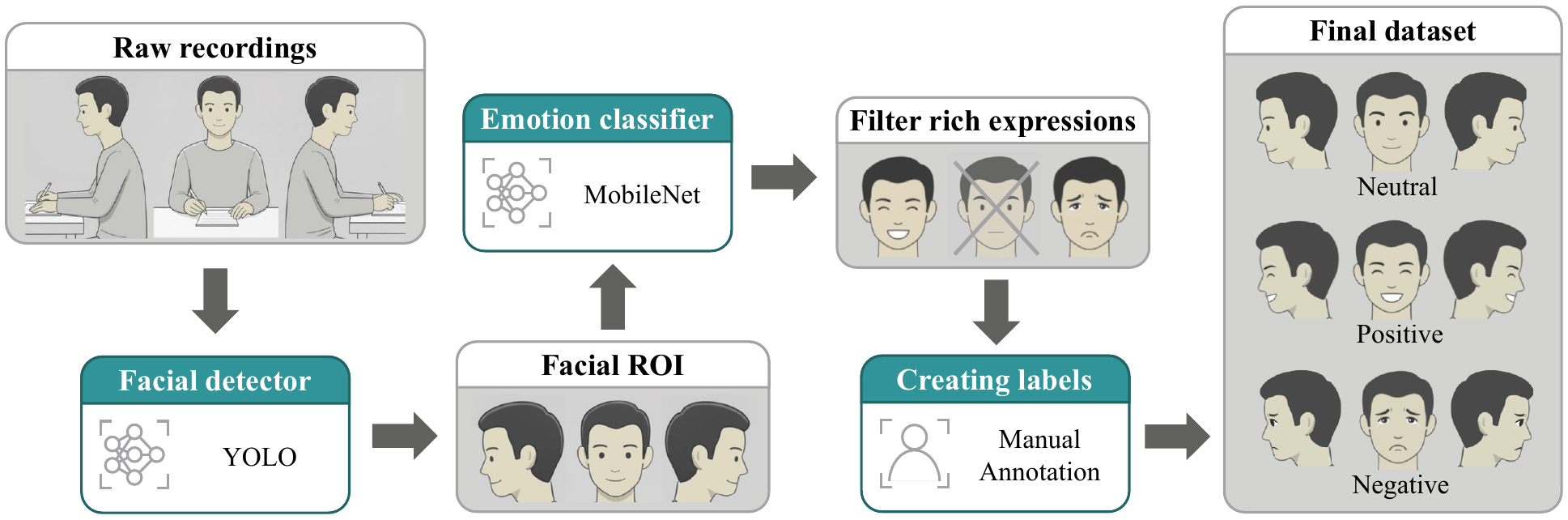}}
	\caption{Pipeline of RGB‐D data processing for constructing the multi‐view facial emotion recognition dataset. The procedure includes raw video acquisition, face detection, ROI extraction, emotion classification, filtering of expressive frames, manual labeling, and final dataset generation.}
	\label{fig:collect_face}
\end{figure*}

The multi-view facial expression subset is constructed through three steps:
(1) face detection using YOLOv8 pretrained on COCO and fine-tuned on WiderFace;
(2) frame-level pre-screening using MobileNetV2 trained on FER2013+;
(3) manual verification by 10 annotators to review the screened frames, remove duplicates, and finally assign three emotion labels (positive, neutral, and negative).
The overall pipeline is illustrated in Fig.~\ref{fig:collect_face}.  

The experimental settings are as follows:

\begin{itemize}
	\item \textbf{Intra-subject EEG experiments:} for each participant, the first 10 clips are used for training and the last 6 clips for testing to prevent temporal leakage;
	\item \textbf{Cross-subject EEG experiments:} we adopt a leave-one-subject-out protocol, alternating each participant as the target domain;
	\item \textbf{Cardiac-signal and multimodal physiological experiments:} we include 17 participants with complete physiological recordings (excluding one participant whose ECG exhibited substantial contact noise);
	\item \textbf{Multi-view visual experiments:} five-fold cross-validation is used to evaluate the stability of tri-view facial emotion recognition.
\end{itemize}

For classification, SVM and KNN are used for physiological-signal tasks, while MobileNetV2 is used for visual tasks. For cross-subject EEG experiments, we additionally evaluate representative domain adaptation models, including DAN, DANN, MS-MDA, NSAL-DGAT, and SSAA-MSDA. Hyperparameters for SVM and KNN are determined via systematic grid search (see Table~\ref{tab:SVM_KNN_param}).

\begin{table}[h]
	\centering
	\caption{Parameter settings and search ranges for SVM and KNN classifiers used in single-subject EEG emotion recognition experiments.}
	\setlength{\tabcolsep}{3pt} % 减小列间距
	\begin{tabular}{cccc}
		\toprule
		Classifier	&	Parameter	&	Value Range\\
		\midrule
		\multirow{2}{*}{SVM}	&	$C$	&	[0.0001, 0.0003, 0.001, 0.003, 0.01, 0.1, 0.3, 1.5, 10]	\\
		&	$\gamma$	&	[0.0001, 0.001, 0.003, 0.01, 0.03, 0.1, 0.5, 1, 5, 10]	\\
		\cmidrule(lr){1-3}
		\multirow{2}{*}{KNN}	&	$n$	&	[3, 5, 7, 9, 11]	\\
		&$p$	&	[1, 2]	\\
		\bottomrule
	\end{tabular}
	\label{tab:SVM_KNN_param}
\end{table}

All experiments are conducted under the same computing environment (Intel Xeon Gold 5218 CPU, 64,GB RAM, NVIDIA RTX 4090 GPU, and Python~3.10). To ensure fair comparisons, we adopt unified data-splitting protocols and evaluation metrics across all experiments.

\subsection{EEG-based Experiments}

\subsubsection{Intra-subject EEG Recognition}

In this experiment, we focus on the discriminative capability of EEG under different emotion-annotation perspectives by comparing emotion classification performance trained with stimulus-based labels versus cognitive labels. This design aims to reveal neural-level differences between externally elicited emotional responses and individuals’ subjective experiences.

The results (Table~\ref{tab:single_sub}) show that models trained with stimulus-based labels significantly outperform those trained with cognitive labels. This suggests that EEG responses to externally elicited emotions are more consistent, whereas cognitive labels are more affected by inter-individual differences, resulting in reduced separability.

To further quantify how label consistency between the two annotation perspectives affects EEG stability, we construct a “label-consistent subset” that retains only samples whose stimulus-based and cognitive labels are identical. Let $y_i^{\text{mm}}$ and $y_i^{\text{cogn}}$ denote the stimulus-based and cognitive labels of the $i$-th sample, respectively; the consistency ratio is defined as:

\begin{equation}
	P_{\text{Cons}} =
	\frac{1}{N}\sum_{i=1}^{N}
	\mathbb{I}(y_i^{\text{cogn}} = y_i^{\text{stim}}),
\end{equation}

The classification accuracy on the consistent subset is defined as:

\begin{equation}
	Acc_{\text{Cons}} =
	\frac{
		\sum_{i=1}^{N}
		\mathbb{I}(y_i^{\text{cogn}} = y_i^{\text{stim}})
		\mathbb{I}(\hat{y}_i = y_i^{\text{cogn}})
	}{
		\sum_{i=1}^{N}
		\mathbb{I}(y_i^{\text{cogn}} = y_i^{\text{stim}})
	}.
\end{equation}

where $\mathbb{I}(\cdot)$ denotes the indicator function and $\hat{y}_i$ is the predicted label.

The results (Table~\ref{tab:congn_consistency}) indicate that EEG classification performance improves markedly when the two label types agree, suggesting that cross-perspective label consistency reflects the stability of neural activity. In contrast, inconsistent samples often correspond to emotion regulation processes or ambiguous affective states. This phenomenon provides neural-level empirical support for the hierarchical separation between “perception” and “experience” in emotion.

\begin{table}[h]
	\centering
	\caption{Single-subject EEG emotion classification results under two labeling paradigms. “Multimedia” refers to the emotion categories assigned to the external stimulus clips, whereas “Cognitive” denotes self-reported affective states after viewing. The figure presents mean accuracy and standard deviation, highlighting the discrepancy between stimulus-driven and self-perceived emotional responses.}
	
	\begin{tabular}{cccccc}
		\toprule
		Task	&	Classifier	&	Label Set	&	Acc.	&	Std.\\
		\midrule
		\multirow{4}{*}{3-class}	
		&	\multirow{2}{*}{SVM}	
		&	Multimedia	&	$\mathbf{0.888}$	&	0.113	\\
		&	&	Cognitive	&	0.693	&	0.158	\\
		\cmidrule(lr){2-5}
		&	\multirow{2}{*}{KNN}	
		&	Multimedia	&	$\mathbf{0.894}$	&	0.153	\\
		&	&	Cognitive	&	0.699	&	0.179	\\
		\cmidrule(lr){1-5}
		\multirow{4}{*}{4-class}	
		&	\multirow{2}{*}{SVM}	
		&	Multimedia	&	$\mathbf{0.908}$	&	0.190	\\
		&	&	Cognitive	&	0.407	&	0.154	\\
		\cmidrule(lr){2-5}
		&	\multirow{2}{*}{KNN}	
		&	Multimedia	&	$\mathbf{0.898}$	&	0.177	\\
		&	&	Cognitive	&	0.419	&	0.140	\\
		\bottomrule
	\end{tabular}
	\label{tab:single_sub}
\end{table}

\begin{table}[h]
	\centering
	\caption{Classification performance of single-subject EEG data when multimedia and cognitive labels are consistent, reporting label consistency rate and accuracy.}
	\begin{tabular}{ccccc}
		\toprule
		Task & Label Set & Consistency Rate & Acc. \\
		\midrule
		\multirow{2}{*}{3-class} 
		& Cognitive (full set) & 1 & 0.693 \\
		& Cognitive(subset) & 0.889 & \textbf{0.727} \\
		\cmidrule(lr){1-4}
		\multirow{2}{*}{4-class} 
		& Cognitive (full set) & 1 & 0.407 \\
		& Cognitive (subset) & 0.889 & \textbf{0.479} \\
		\bottomrule
	\end{tabular}
	\label{tab:congn_consistency}
\end{table}

\begin{table}[h]
	\centering
	\caption{Comparison of single-subject EEG emotion recognition performance between MAD and benchmark datasets under identical experimental settings.}
	\begin{tabular}{cccccc}
		\toprule
		\multirow{2}{*}{Task}	&		\multirow{2}{*}{Dataset}		&	\multirow{2}{*}{Label Set}	&	\multicolumn{2}{c}{Acc.}\\
		\cmidrule(lr){4-5}
		&	&	&	SVM	&	KNN	\\
		\midrule
		\multirow{7}{*}{3-class}	
		&	SEED~\cite{zheng2015investigating}	&	Multimedia	&	0.745	&	0.738	\\
		&	DEAP(Valence)	&	Cognition	&	0.501	&	0.507	\\
		&	DEAP(Arousal)	&	Cognition&	0.672	&	0.620	\\
		&	Dreamer(Valence)	&	Cognition&	0.473	&	0.447	\\
		&	Dreamer(Arousal)	&	Cognition&	0.429	&	0.404	\\
		&	MAD(cogn)	&	Cognition&	0.693	&	0.699	\\
		&	MAD(mm)	&	Multimedia&	0.888	&	0.894	\\
		\cmidrule(lr){1-5}
		\multirow{5}{*}{4-class}	
		&	SEED $\mathrm{IV}$~\cite{zheng2018emotionmeter}	&	Multimedia	&	0.637	&	0.623	\\
		&	DEAP~\cite{koelstra2011deap}	&	Cognition&	0.526	&	0.547	\\
		&	Dreamer~\cite{katsigiannis2017dreamer}	&	Cognition	&	0.413	&	0.437	\\
		&	MAD(cogn)	&	Cognition&	0.407	&	0.417	\\
		&	MAD(mm)	&	Multimedia&	0.908	&	0.898	\\
		\bottomrule
	\end{tabular}
	\label{tab:self_other}
\end{table}

To validate the reliability and generalizability of MAD, we further compare it with several widely used EEG emotion datasets, including SEED, SEED-IV, DEAP, and DREAMER~\cite{katsigiannis2017dreamer}. All comparison experiments adopt identical feature extraction procedures and classifier configurations to ensure comparability.

Regarding annotation paradigms, SEED and SEED-IV use stimulus-based emotion labels, whereas DEAP and DREAMER are annotated based on participants’ self-reported assessments. In terms of task settings, SEED provides three emotion classes and SEED-IV extends this to four classes; DEAP and DREAMER are labeled in the valence–arousal space and further define three-class subtasks along the valence and arousal dimensions.
The comparison results are summarized in Table~\ref{tab:self_other}. Under stimulus-based supervision, MAD achieves performance comparable to mainstream datasets and slightly better on some tasks; under self-report supervision, MAD exhibits trends that are consistent and stable relative to datasets with similar labeling schemes. These results indicate that MAD provides reliable data quality and reasonable task difficulty under different annotation paradigms, making it a benchmark dataset with strong generalization potential for EEG-based emotion recognition.

\subsubsection{Cross-subject EEG Emotion Recognition}

Cross-subject EEG emotion recognition faces substantial challenges due to inter-individual variability. Differences in neural activity patterns and physiological characteristics across participants can cause significant distribution shifts in EEG features, thereby limiting generalization to unseen subjects. In recent years, \textit{domain adaptation} (DA) methods have been widely used to mitigate this issue. The central idea is to reduce distribution discrepancies between the source domain (training subjects) and the target domain (test subject) via feature alignment, thus improving cross-individual transfer performance.

In this experiment, all cross-subject EEG recognition tasks are conducted under \textbf{stimulus-based labels} to ensure that emotion categories share a consistent external reference across subjects. To systematically evaluate the applicability of MAD and its compatibility with representative algorithms in cross-subject transfer learning, we compare five representative DA methods, including DAN, DANN, MS-MDA, NSAL-DGAT, and SSAA-MSDA, covering major paradigms such as MMD-based kernel alignment, adversarial learning, graph attention mechanisms, and multi-source domain adaptation. All methods are implemented under unified preprocessing, feature extraction, and training protocols to ensure fair and comparable results.

As shown in Table~\ref{tab:cross_other}, none of the five DA methods exhibits performance collapse on MAD; results remain stable and the performance differences among methods are relatively small. This suggests that MAD presents a cross-subject variability structure that is “realistic yet learnable”: it preserves natural individual differences, providing room for DA optimization, while maintaining sufficient signal quality and emotional separability to support stable cross-domain transfer.

In summary, this experiment confirms that MAD can serve as a standardized benchmark for cross-subject EEG emotion recognition and domain adaptation research, providing reliable data support for personalized emotion modeling and multi-source transfer learning.

\begin{table}[h]
	\centering
	\caption{Performance of representative domain adaptation algorithms for cross-subject EEG emotion recognition on MAD.}
	\begin{tabular}{ccccc}
		\toprule
		Task	&	Method	&	Acc.	&	Std.\\
		\midrule
		\multirow{5}{*}{3-class}	
		&	DAN	\cite{li2019domain}&	0.657	&	0.042	\\
		&	DANN~\cite{tzeng2019deep}	&	0.642	&	0.036	\\
		&	MS-MDA~\cite{chen2021ms}	&	0.617	&	0.053	\\
		&	NSAL-DGAT~\cite{yang2025exploiting}	&	0.542	&	0.045	\\
		&	SSAA-MSDA~\cite{yang2024spectral}	&	0.613	&	0.046	\\
		\cmidrule(lr){1-4}
		\multirow{5}{*}{4-class}	
		&	DAN	&	0.654	&	0.057	\\
		&	DANN	&	0.679	&	0.090	\\
		&	MS-MDA	&	0.566	&	0.096	\\
		&	NSAL-DGAT	&	0.526	&	0.096	\\
		&	SSAA-MSDA	&	0.597	&	0.087	\\
		\bottomrule
	\end{tabular}
	\label{tab:cross_other}
\end{table}

\subsection{Emotion Recognition with ECG, PPG, and BCG}

Electrocardiography (ECG), photoplethysmography (PPG), and ballistocardiography (BCG) all originate from the same cardiovascular activity process, characterizing cardiac electrical excitation, peripheral blood volume changes, and mechanical vibrations induced by cardiac beats, respectively. Although their sensing mechanisms and measurement modalities differ, their temporal rhythms are jointly driven by the cardiac systole–diastole cycle and should therefore exhibit highly consistent temporal patterns in principle. The high-precision synchronization framework of MAD provides a reliable foundation for systematic cross-modal consistency analysis among these cardiac-related signals.

\begin{figure}[!t]
	\centering
	\subfloat[]{\includegraphics[width=3.5in]{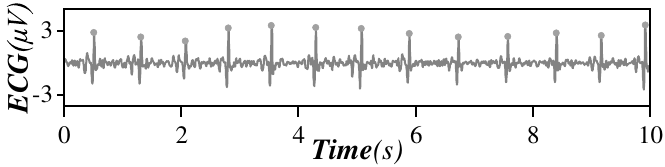}%
		\label{fig:signal-ecg}}
	\hfil
	
	\subfloat[]{\includegraphics[width=3.5in]{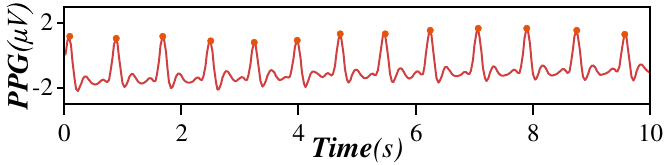}%
		\label{fig:signal-ppg}}
	
	\subfloat[]{\includegraphics[width=3.5in]{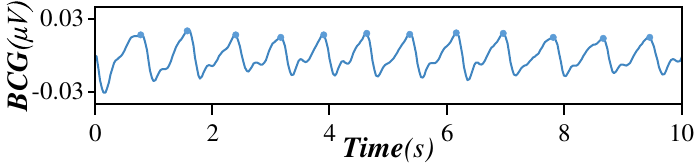}%
		\label{fig:signal-bcg}}
	\caption{Representative waveforms of ECG, PPG, and BCG signals with annotated peaks. Subplots (a)–(c) show ECG, PPG, and BCG signals, respectively, with R-peaks in ECG, Systolic-peaks in PPG, and J-peaks in BCG indicated.The 10-second segments were selected from 30-second synchronous recordings to analyze heart cycle consistency.}
	\label{fig:signal-ecgecgecg}
\end{figure}

In this experiment, all emotion recognition and consistency analyses are conducted under \textbf{stimulus-based labels} to ensure a unified external reference across subjects. To evaluate rhythm-level correspondence among the three modalities, we extract continuous 10-second segments from synchronized 30,s multimodal recordings for peak annotation. Figure~\ref{fig:signal-ecgecgecg} shows representative waveforms of ECG, PPG, and BCG, with ECG R-peaks, PPG systolic peaks, and BCG J-peaks annotated, respectively. Despite differences in amplitude morphology and noise characteristics, the peak sequences exhibit clear temporal synchrony, visually confirming strong rhythm-level consistency across the three cardiac signals.

\begin{figure}[!t]
	\centering
	\includegraphics[width=.98\linewidth]{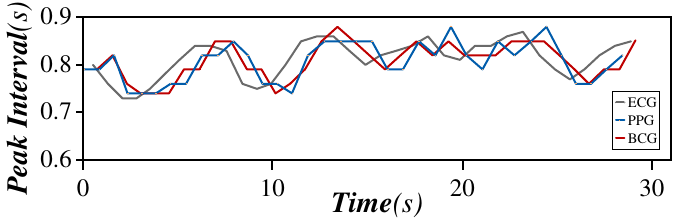}
	\caption{Comparison of peak-to-peak intervals for ECG, PPG, and BCG signals. The plot shows R-R intervals in ECG, Systolic-Systolic intervals in PPG, and J-J intervals in BCG, illustrating consistent heart cycles across modalities.}
	\label{fig:signal-diff}
\end{figure}

Furthermore, Fig.~\ref{fig:signal-diff} presents the inter-beat interval sequences of the three modalities over the full 30,s window. The results show nearly identical fluctuation trends and highly overlapping heart-rate variation curves, indicating strong synchrony in how these modalities respond to autonomic regulation.

For emotion recognition, we extract the same time-domain and nonlinear features from ECG, PPG, and BCG, and evaluate both three-class (positive, neutral, negative) and four-class (happy, sad, fearful, neutral) tasks. Since one participant’s ECG recordings contained substantial contact noise, a total of 17 participants are included in this analysis.

\begin{figure}[h]
	\centering
	\subfloat[]{\includegraphics[width=2.3in]{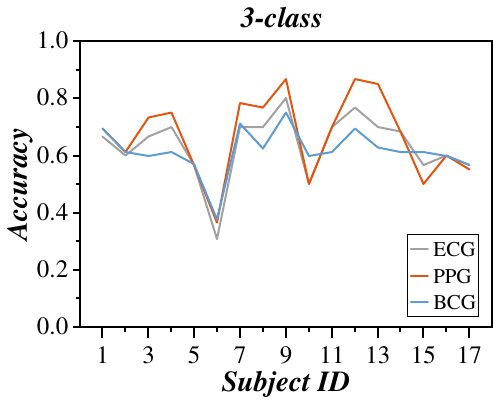}%
		\label{fig:ecg_sub3}}
	\hfil
	
	\subfloat[]{\includegraphics[width=2.3in]{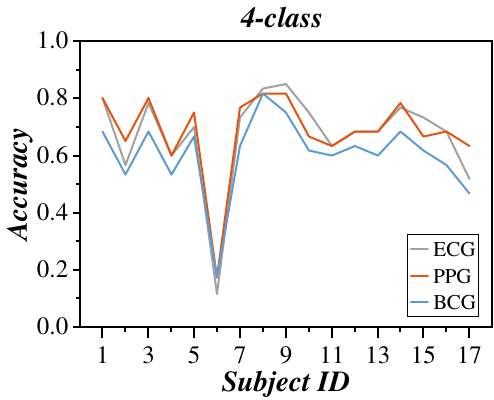}%
		\label{fig:ecg_sub4}}
	
	\caption{Classification accuracy of ECG, PPG, and BCG signals in single-subject emotion recognition experiments. Subplot (a) shows three-class results, and subplot (b) shows four-class results across 17 subjects, with each line representing one signal modality.}
	\label{fig:ecg_result}
\end{figure}

As shown in Fig.~\ref{fig:ecg_result}, the three cardiac-related signals exhibit highly consistent trends in classification performance, with most accuracies distributed in the 50--80\% range. Notably, BCG, as a non-contact measurement modality, achieves performance comparable to contact-based ECG and PPG, suggesting its potential for emotion recognition in wearable or passive sensing scenarios.

Overall, MAD’s high-precision synchronization ensures rhythm-level consistency among the three cardiac modalities, and the consistency in their classification trends further supports the stable role of cardiovascular dynamics in emotion regulation. In future work, the RGB-D videos in MAD may enable remote PPG (rPPG) reconstruction, facilitating exploration of vision–physiology fusion and the possibility of fully non-contact emotion recognition.

\subsection{Multimodal Fusion Experiments}

This experiment systematically evaluates the effectiveness of multimodal physiological feature fusion for emotion recognition, including EEG, ECG, EOG, and EMG. All emotion recognition tasks are supervised by \textbf{stimulus-based labels} to ensure a unified and stable emotion reference across modalities and subjects. Since ECG effectively characterizes heart-rate rhythms and autonomic activity, and PPG/BCG are highly correlated with ECG in physiological information, we exclude PPG and BCG from fusion analysis to avoid redundancy and repeated information. Ultimately, 17 participants with complete multimodal recordings are included.

\begin{figure}[!t]
	\centering
	\subfloat[]{\includegraphics[width=2.7in]{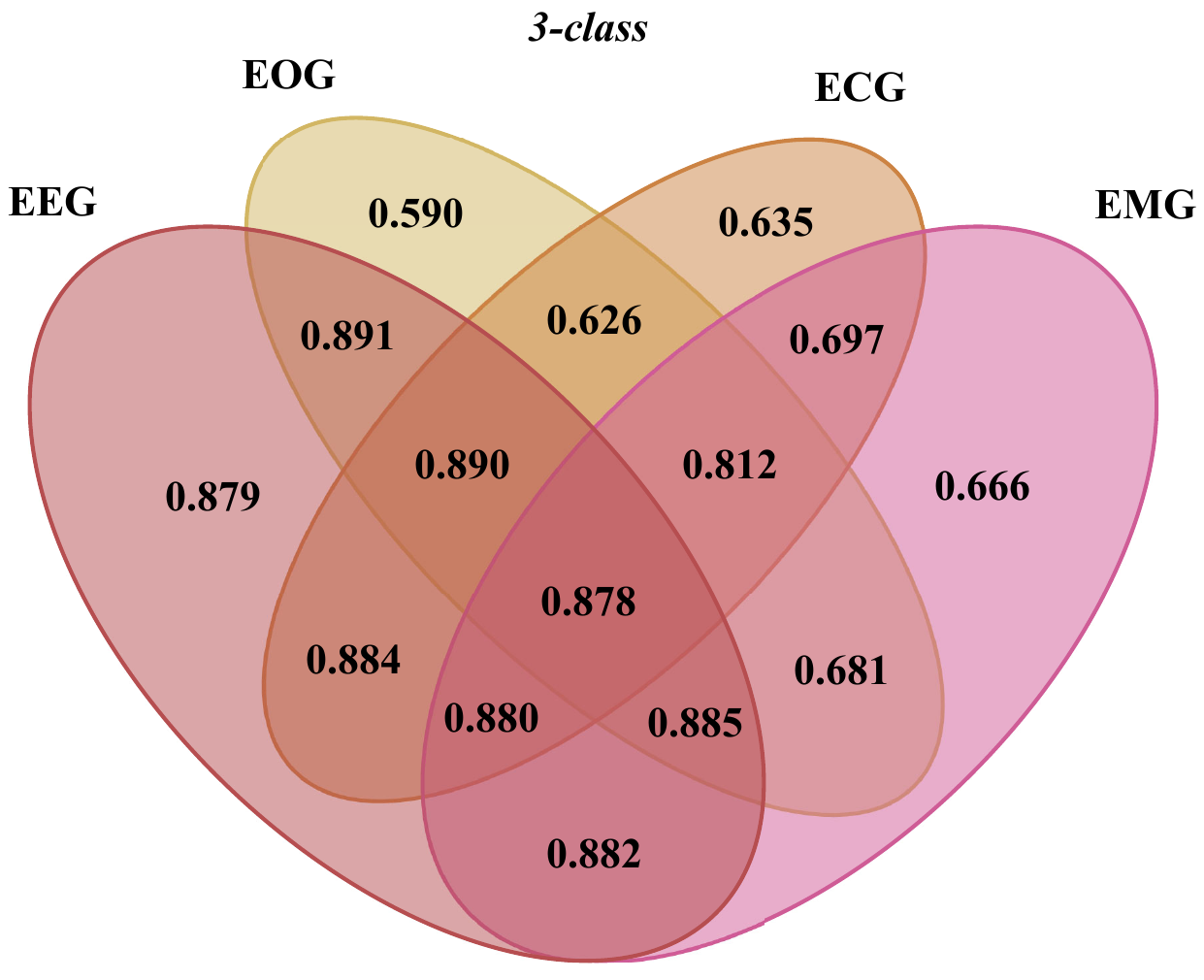}%
		\label{fig:mm_venny_3}}
	\hfil
	
	\subfloat[]{\includegraphics[width=2.7in]{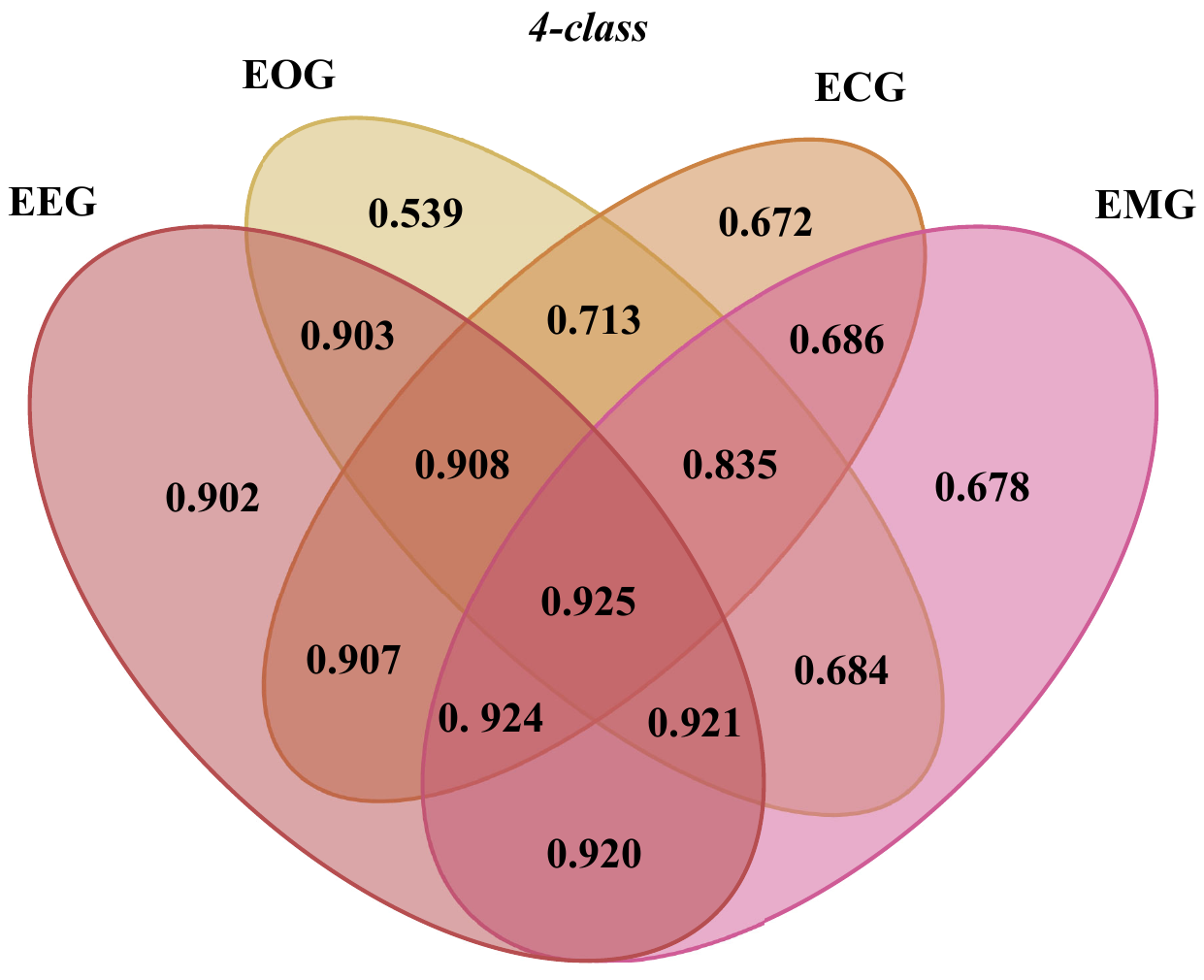}%
		\label{fig:mm_venny_4}}
	\caption{Venn diagrams showing the classification performance of individual and combined modalities (EEG, ECG, EOG, EMG) in multi-subject emotion recognition. Subplot (a) corresponds to three-class and subplot (b) to four-class tasks, illustrating overlapping contributions of different signal combinations.}
	\label{fig:mm_venny}
\end{figure}

The results are shown in Fig.~\ref{fig:mm_venny}. Under unimodal settings, EEG consistently achieves the best performance, further confirming its central role in non-invasive emotion recognition. In contrast, peripheral modalities (ECG, EOG, and EMG) yield relatively limited performance when used alone; however, their combinations (e.g., EOG+ECG+EMG) significantly outperform any single peripheral modality, reflecting complementarity among different physiological regulation systems during emotional responses.

It is worth noting that even fusing multiple peripheral modalities does not surpass EEG alone, indicating that the primary discriminative information for emotion recognition remains dominated by central neural activity. When EEG is further fused with peripheral modalities (e.g., EEG+ECG or EEG+EOG), additional performance gains are observed, suggesting that peripheral signals, while not dominant, can complement EEG representations under certain affective intensities or arousal levels.

Overall, this experiment validates a hierarchical distribution of emotion-related physiological information across modalities: EEG provides the most discriminative central representation, while peripheral physiological signals reflect autonomic and behavioral regulation states and serve as auxiliary enhancements for emotion recognition. MAD demonstrates strong experimental consistency in multimodal synchronization, modality complementarity, and hierarchical fusion analysis, providing a reliable platform for systematic research on multimodal physiological emotion modeling.

\subsection{Multi-view Facial Emotion Experiments}

Facial expressions provide the most intuitive and readily perceivable channel of overt emotional behavior. However, in real-world applications, head-pose variations, viewpoint shifts, and partial occlusions can substantially degrade the stability and generalization of single-view facial emotion recognition models. Traditional single-view approaches often rely on frontal-view local cues (e.g., eye and mouth regions); when these regions become partially invisible due to pose changes, performance can deteriorate noticeably. Incorporating multi-view information can therefore enhance robustness in facial emotion recognition and also lays a foundation for subsequent vision–physiology cross-modal emotion modeling.

Based on the tri-view RGB-D facial videos in MAD, this experiment systematically evaluates the role of multi-view information in emotion recognition. All visual experiments use \textbf{expression labels}, which are assigned by human annotators according to observable facial behaviors to reflect the overt-expression level of emotion. To progressively enhance multi-view learning capability, we design four experimental configurations:

\begin{itemize}
	\item \textbf{Baseline:} directly test a model pretrained on FER2013plus on MAD without fine-tuning to evaluate cross-dataset transferability;
	\item \textbf{Single-view fine-tuning:} fine-tune the model using frontal-view images only to assess adaptation to MAD under a single-view setting;
	\item \textbf{Multi-view fine-tuning:} jointly train with frontal and bilateral-view data to analyze how multi-view samples mitigate pose bias and improve robustness;
	\item \textbf{Multi-view feature alignment:} on top of multi-view fine-tuning, introduce a contrastive learning mechanism to aggregate synchronized tri-view frame features and enforce cross-view consistency of emotion representations in latent space using a cosine-similarity-based contrastive loss, thereby learning view-invariant emotion representations.
\end{itemize}

All experiments adopt a three-class task (positive, neutral, negative). The original labels in FER2013plus are mapped to the same three-class taxonomy to ensure annotation consistency between pretraining and downstream evaluation. Each frame in MAD is independently scored by 10 annotators, and the final label is obtained by averaging the scores and discretizing the result. We use cross-subject five-fold cross-validation and report classification accuracy and cross-entropy loss as evaluation metrics.

The experimental results are summarized in Table~\ref{tab:multi-view-result}. Under the baseline setting, recognition accuracies are low across the three views (0.428–0.554), reflecting the difficulty of direct cross-dataset transfer in facial emotion recognition. After single-view fine-tuning, the frontal-view accuracy increases substantially to 0.835, indicating that the model can effectively learn expression distributions in MAD. With multi-view fine-tuning, performance on side views improves markedly and becomes more balanced across views (0.734–0.786), validating the effectiveness of multi-view training in mitigating pose bias.
Furthermore, incorporating cross-view contrastive learning achieves the best results (frontal: 0.845, side views: 0.818, fusion: 0.827). These results suggest that contrastive constraints can effectively bring representations of the same emotion across different viewpoints closer in feature space, encouraging the model to learn view-invariant emotional semantics and thereby achieving more stable and consistent recognition.

Overall, the analysis indicates that: (1) multi-view training significantly alleviates performance degradation caused by pose variations; (2) cross-view contrastive learning improves feature-space consistency and robustness; and (3) the tri-view design of MAD provides a solid data foundation for cross-view facial emotion recognition as well as vision–physiology emotion fusion research.

\begin{table}[h]
	\centering
	\caption{Comparison of Single-view and Multi-view Fine-tuning Strategies on the MAD Dataset for Three-class Facial Emotion Recognition (Positive, Neutral, Negative)}
	\setlength{\tabcolsep}{3pt} % 减小列间距
	\begin{tabular}{cccc}
		\toprule
		\multirow{2}{*}{Training Strategy}	&	 \multicolumn{3}{c}{Acc.}	\\
		\cmidrule(lr){2-4}
		&	Frontal View	&	Side Views	&	All Views\\
		\midrule
		Baseline (Pre-trained only)	&	0.554	&	0.428	&	0.457	\\
		Single-view Fine-tuning	&	0.835	&	0.682	&	0.735	\\
		Multi-view Fine-tuning	&	0.734	&	0.786	&	0.767	\\
		Multi-view Aggregation	&	\textbf{0.845}	&	\textbf{0.818}	&	\textbf{0.827}	\\
		\bottomrule
	\end{tabular}
	\label{tab:multi-view-result}
\end{table}

\section{Discussion}

By synchronously recording neural, physiological, and behavioral signals, MAD provides a systematic cross-modal and cross-level observation perspective for emotional processes. Its acquisition and annotation procedures explicitly distinguish among stimulus elicitation, subjective experience, and overt expression. As a result, MAD can serve not only as a benchmark for algorithm evaluation but also as a resource for mechanism-oriented analyses that align with theories in psychology and neuroscience.

\subsection{Differences Between Stimulus-Based and Cognitive Labels}

The intra-subject EEG experiments show that models trained with stimulus-based labels generally outperform those trained with cognitive labels. This observation is consistent with fundamental insights in emotion research: stimulus labels are externally defined and tend to be more consistent across repeated trials, whereas cognitive labels are influenced by individuals’ expressive ability, attention allocation, and emotion regulation strategies, leading to larger subjective variability. Meanwhile, EEG encoding of implicit emotional experiences may be more distributed and complex, making it difficult for models to learn stable patterns directly from cognitive labels. Therefore, retaining both stimulus-based and cognitive labels is meaningful: they respectively capture externally elicited emotions and internal subjective experiences, providing complementary information for multi-perspective emotion analysis.

\subsection{Cross-Subject Variability and the Feasibility of Adaptive Learning}

EEG signals commonly exhibit substantial inter-individual variability. On MAD, multiple domain adaptation methods achieve stable cross-subject performance, indicating that the inter-subject differences captured in the dataset are realistic yet remain within a learnable range. This property allows MAD to reflect genuine cross-individual variability without causing model training to fail due to unstable signals, making it suitable as an evaluation platform for cross-subject transfer learning and personalized emotion modeling.

\subsection{Correlation Among Cardiac-Related Physiological Signals}

Although ECG, PPG, and BCG differ in sensing mechanisms, they exhibit highly consistent heart-rate-variability rhythms in MAD, validating the reliability of the acquisition and synchronization system and indicating that different cardiac modalities can reflect consistent autonomic regulation states. This result further supports the potential substitutability of PPG and BCG for emotion recognition, especially in wearable or low-contact scenarios with practical relevance. In addition, the synchronized RGB-D videos in MAD provide a foundation for future exploration of vision-based non-contact cardiac signal reconstruction (e.g., rPPG).

\subsection{Complementarity of Multimodal Physiological Signals}

The multimodal fusion experiments indicate that EEG remains the most discriminative single modality, while peripheral signals such as ECG, EOG, and EMG can complement EEG to some extent and improve overall robustness. This suggests that emotional states involve coordinated responses across multiple physiological systems, and different modalities offer advantages along distinct regulation pathways and time scales. Future work may further exploit complementary information among physiological systems through more refined fusion strategies (e.g., adaptive weighting or cross-modal alignment).

\subsection{Significance of Multi-view Facial Expression Learning}

The multi-view facial expression experiments demonstrate that incorporating multi-view training and cross-view contrastive constraints can effectively improve robustness to pose variations and facilitate learning of view-invariant emotion representations. This result not only validates the effectiveness of MAD’s tri-view design but also provides important support for future studies on cross-modal consistency between visual expressions and physiological responses.

\subsection{Limitations and Future Work}

Despite providing a rich multimodal foundation for affective computing, MAD is still primarily collected in controlled laboratory settings with a limited participant scale, and it does not yet cover complex real-world contexts and noise factors. Future work will expand the participant pool and explore continuous emotion data collection in more natural scenarios using wearable or mobile platforms. Moreover, MAD’s multi-level annotation scheme enables the development of hierarchical and interpretable emotion models, which may facilitate deeper analysis of the relationships among stimulus elicitation, subjective experience, and behavioral expression.

\section{Conclusion}

In this work, we construct and release the Multimodal Affection Dataset (MAD), a large-scale multimodal dataset for affective computing with high synchronization precision. MAD integrates six physiological modalities (EEG, ECG, EOG, EMG, PPG, and BCG) with tri-view RGB-D facial videos, and introduces a three-level emotion annotation framework at the stimulus, cognitive, and expression levels. This dataset provides an unprecedented experimental foundation for investigating relationships among emotion elicitation, subjective experience, and behavioral expression.

Through systematic benchmark experiments, we validate the value of MAD from multiple perspectives:

\begin{itemize}
	\item We reveal neural-level differences between externally elicited emotions and subjective experiences;
	\item We show that MAD presents stable cross-subject transfer difficulty, supporting research on domain adaptation;
	\item We verify the consistency of cardiac-related signals in autonomic regulation and their potential substitutability;
	\item We demonstrate the potential of multimodal fusion for improving robustness in emotion recognition;
	\item We demonstrate that multi-view contrastive learning enables pose-invariant facial emotion recognition.
\end{itemize}

Overall, MAD establishes a unified emotion research platform spanning neural, physiological, and behavioral levels. It can serve both as a standardized benchmark for algorithm development and as an experimental resource for exploring emotion mechanisms.

In future work, we will further expand the dataset scale and scenario diversity, and incorporate foundation models and multimodal pretraining techniques to advance interpretable, generalizable, and deployable affective computing systems. MAD aims to serve as a bridge between fundamental theory and practical applications, accelerating progress in human–machine collaboration and emotion understanding technologies.

\section*{Declarations}

\subsection*{Author Contributions}
S.G.: Led the study design, data acquisition strategy, multimodal experiments, and manuscript preparation.
Y.Q. and W.Z.: Contributed to experimental procedures and data collection.
B.L. and Y.W.: Provided equipment support and technical assistance.
G.S.: Oversaw the project, coordinated research activities, and provided primary academic guidance.

\subsection*{Acknowledgements}
The authors would like to express their sincere appreciation to all volunteers for their participation in the MAD data collection, to the annotators for their dedicated efforts in labeling the dataset, and to the members of the Biomedical Signal Processing Laboratory for their invaluable technical assistance. The authors also gratefully acknowledge the constructive support and contributions from all coauthors throughout the development of this work.

\subsection*{Funding}
The authors received no specific funding for this work.

\subsection*{Conflict of Interest}
The authors declare that they have no conflict of interest.

\bibliographystyle{IEEEtran}
\bibliography{affect_bibliography}

\end{document}